\documentstyle[twoside,A4]{article}
\advance\topmargin .5in

\begin{document}
\thispagestyle{plain}
\title{\marginpar{\vspace{-1in}\hspace{-1in}\small KFT U\L\ 5/93}%
TWO-DIMENSIONAL QUANTUM POINCAR\'E GROUP\thanks{Supported by KBN
grant No. 2 0218 91 01}
\thanks{Talk given by J.R. during the ``Conference on Quantum Topology'',
Manhattan, Kansas, 24--28 March 1993.}}

\author{JAKUB REMBIELI\'NSKI\thanks{e-mail: {\tt jaremb@plunlo51.bitnet}}\\
{\em Department of Mathematical Physics, University of {\L}\'od\'z}\\
{\em ul.\ Pomorska 149/153, 90--236 {\L}\'od\'z, Poland}\\[0.3cm]
and\\[0.3cm]
WAC\L AW TYBOR\thanks{e-mail: {\tt watybor@plunlo51.bitnet}}\\
{\em Department of Theoretical Physics, University of {\L}\'od\'z}\\
{\em ul.\ Pomorska 149/153, 90--236 {\L}\'od\'z, Poland}}
\date{}

\maketitle

\begin{abstract}
Quantum Poincar\'e-Weyl group in two dimensional quantum Minkowski space-time
is considered and an appriopriate relativistic kinematics is investigated.  It
is claimed that a consistent approach to the above questions demands a kind of
a ``quantum geometry'' in the $q$-deformed space-time.
\end{abstract}

\section{Introduction}
Recently, in a number of papers
\cite{aref1,remb4,remb2,remb3,brzez1,ubr1,schw1,cab1} it was considered a
variety of models realising the ideas of quantization of space-time and
phase-space by means of the non-commutative geometry and quantum group theory
methods.  In particular in \cite{aref1} it was considered simple
one-dimensional non-commutative dynamical systems.  A very intriguing feature
of that models is non-commutativity of the inertial mass.  However, the
non-commutativity of the coordinates with inertial mass is no so satisfactory
property because of generically different physiacal meaning of the coordinates
(geometrical objects) and the inertial mass (dynamical object).

In this paper we try to solve simultanously two problems: (a) the indicated
question of the non-commutativity of the inertial mass, (b) the rescaling of
the $q$-Lorentz invaraints related to the fact that quantum determinant does
not belong to the center of the $q$-Poincar\'e group \cite{remb4}.  This can
be done by introducing in the $q$-deformed Minkowski space-time a kind of the
``quantum geometry''by means of the notion of ``quantum metrics''.  In the
classical limit the standard Lobatshevsky geometry is recovered.  Moreover,
in a contrast to the \cite{remb4} we give a geometric interpretation of momenta
as the translation generators defined via corresponding quantum Cartan-Maurer
forms.

\section{Quantum Minkowski Space-Time and \newline
the Quantum Poincar\'e-Weyl Group}
We assume that the $q$-deformed Minkowski space-time is a {\em real form\/} of
the Manin's plane.  Consequently the $q$-Minkowski space-time is generated by
generators $x_\pm$ satysfying
\begin{equation}
x_+x_-=qx_-x_+,
\end{equation}
with $x^*_\pm=x_\pm$ and $|q|=1$. We will interprete $x_\pm$ as light-cone
coordinates.

The resulting Poincar\'e-Weyl group is a subgroup of the $IGL(2)_{q,s,\mu}$
quantum group introduced in \cite{remb1} with the co-module action of the form
\begin{equation}
\pmatrix{x'_+\cr x'_-\cr 1}=\delta\pmatrix{x_+\cr x_-\cr 1}
=\pmatrix{\sigma\omega&0&u_+\cr0&\sigma\omega^{-1}&u_-\cr0&0&1}
\otimes\pmatrix{x_+\cr x_-\cr 1},
\end{equation}
where
$\sigma^2
=\det_q\pmatrix{\sigma\omega&0&u_+\cr0&\sigma\omega^{-1}&u_-\cr0&0&1}$ and
$\sigma^*=\sigma$, $\omega^*=\omega$, $u^*_\pm=u_\pm$.

The generators of the above $q$-Poincar\'e-Weyl group fulfil the following
algebraic rules
\begin{eqnarray}
\omega\sigma&=&\sigma\omega,\\
u_\pm\sigma&=&q^{\pm\frac{1}{2}}\sigma u_\pm,\\
\omega u_\pm&=&q^\frac{1}{2}u_\pm\omega,\\
u_+u_-&=&qu_-u_+.
\end{eqnarray}
We will follow with the standard Woronowicz form of co-product \cite{wor1}.
The antipode and co-unity have the form
\begin{equation}
g^{-1}=\pmatrix
{\sigma^{-1}\omega^{-1}&0&-\sigma^{-1}\omega^{-1}u_+\cr
0&\sigma^{-1}\omega&-\sigma^{-1}\omega u_-\cr
0&0&1};
\end{equation}
\begin{equation}
\epsilon(\sigma)=\epsilon(\omega)=1,\qquad\epsilon(u_\pm)=0.
\end{equation}

\section{Bicovariant Differential Calculus on Quantum Poincar\'e-Weyl Group}
We can formulate a bicovariant differential calculus on the above qauntum group
\cite{wor2} with help of the usual differentials ${\rm d}\sigma$, ${\rm
d}\omega$, ${\rm d}u_\pm$
\begin{eqnarray}
\sigma\,{\rm d}\sigma&=&{\rm d}\sigma\,\sigma,\\
\sigma\,{\rm d}\omega&=&{\rm d}\omega\,\sigma,\\
\omega\,{\rm d}\omega&=&{\rm d}\omega\,\omega,\\
\omega\,{\rm d}\sigma&=&{\rm d}\sigma\,\omega,
\end{eqnarray}
\begin{eqnarray}
u_\pm\,{\rm d}\omega&=&q^{-\frac{1}{2}}{\rm d}\omega\,u_\pm,\\
u_\pm\,{\rm d}\sigma&=&q^{\pm\frac{1}{2}}{\rm d}\sigma\,u_\pm,\\
\omega\,{\rm d}u_\pm&=&q^\frac{1}{2}{\rm d}u_\pm\,\omega,\\
\sigma\,{\rm d}u_\pm&=&q^{\mp\frac{1}{2}}{\rm d}u_\pm\,\sigma,\\
u_\pm\,{\rm d}u_\pm&=&{\rm d}u_\pm\,u_\pm,\\
u_\pm\,{\rm d}u_\mp&=&q^{\pm1}{\rm d}u_\mp\,u_\pm,
\end{eqnarray}
\begin{eqnarray}
{\rm d}u_\pm\,{\rm d}\omega&=&-q^{-\frac{1}{2}}\,{\rm d}\omega\,{\rm d}u_\pm,\\
{\rm d}u_\pm\,{\rm d}\sigma&
                          =&-q^{\pm\frac{1}{2}}\,{\rm d}\sigma\,{\rm d}u_\pm,\\
{\rm d}u_+\,{\rm d}u_-&=&-q\,{\rm d}u_-\,{\rm d}u_+,\\
{\rm d}\omega\,{\rm d}\sigma&=&-{\rm d}\sigma\,{\rm d}\omega,\\
({\rm d}\sigma)^2=({\rm d}\omega)^2&=&({\rm d}u_\pm)^2=0.
\end{eqnarray}

Now we introduce the differential forms
\begin{equation}
\Sigma=\sigma^{-1}\,{\rm d}\sigma,\quad
\Omega=\omega^{-1}\,{\rm d}\omega,\quad
T_\pm=\sigma^{-1}\omega^{\mp1}{\rm d}u_\pm,
\end{equation}
with the following commutation relations
\begin{eqnarray}
\Sigma\Omega&=&-\Omega\Sigma,\\
\Sigma T_\pm&=&-T_\pm\Sigma,\\
\Omega T_\pm&=&-T_\pm\Omega,\\
T_+T_-&=&-q^{-1}T_-T_+.
\end{eqnarray}
Then we can define, via the Cartan-Maurer form
\begin{equation}
g^{-1}\,{\rm d}g\equiv{\rm i}(D\Sigma+K\Omega+P_+T_++P_-T_-)
\end{equation}
the generators of the scaling $D$---dilatation, the Lorentz boost $K$ and
translations $P_\pm$.  Consequently
\begin{eqnarray}
{}[D,K]&=&0,\\
{}[D,P_\pm]&=&-{\rm i}P_\pm,\\
{}[K,P_\pm]&=&\mp{\rm i}P_\pm,\\
{}[P_+,P_-]_{q^{-1}}&=&0.\label{qcommut}
\end{eqnarray}
Here $[P_+,P_-]_{q^{-1}}=P_+P_--q^{-1}P_-P_+$.

\section{Covariant Differential Calculus on the $q$-Min\-kow\-ski Space-Time
and
the $q$-Kine\-mat\-ics}
To construct a kind of relativistic kinematics it is necessary to have a
covariant differential calculus on the $q$-Minkowski space.  Taking into
account the classification of differential calculi for the Manin's plane
\cite{brzez2}, we obtain
\begin{eqnarray}
x_\pm\,{\rm d}x_\pm&=&{\rm d}x_\pm\,x_\pm,\\
x_\pm\,{\rm d}x_\mp&=&q^{\pm1}\,{\rm d}x_\mp\,x_\pm,\\
{\rm d}x_+\,{\rm d}x_-&=&-q\,{\rm d}x_-\,{\rm d}x_+,\\
({\rm d}x_\pm)^2&=&0.
\end{eqnarray}
Now we assume that $x_\pm$ depend on an affine parameter $\tau$, i.e.\
\begin{equation}
x_\pm=x_\pm(\tau),\quad{\rm d}x_\pm=\dot x_\pm\,{\rm d}\tau.
\end{equation}
Consequently, under the existence of the free motion (we admit solutions $\ddot
x_\pm=0$), we obtain
\begin{eqnarray}
x_\pm\dot x_\mp&=&q^{\pm1}\dot x_\mp x_\pm,\\
x_\pm\dot x_\pm&=&\dot x_\pm x_\pm,\\
\dot x_+\dot x_-&=&q\dot x_-\dot x_+.
\end{eqnarray}

Momentum is defined in a standard way. We assume that the inertial mass $m$
commute with $x_\pm$ (i.e.\ $m$ has not any geometrical meaning).
\begin{equation}
m^*=m,\quad\dot m=0,
\end{equation}
\begin{equation}
mx_\pm=x_\pm m.
\end{equation}
Now for a free particle
\begin{equation}
p_\pm=m\dot x_\pm,
\end{equation}
so
\begin{equation}
p^*_\pm=p_\pm,
\end{equation}
\begin{equation}
p_+p_-=qp_-p_+,\label{momenta}
\end{equation}
according to the geometrical interpretation of $p_\pm$.

\section{Quantum Geometry}
The difference between powers of $q$ in the Eq.~(\ref{qcommut}) and
Eq.~(\ref{momenta}) is related to the contravariant nature of $p_\pm=m\dot
x_\pm$ and covariant nature of $P_\pm$.  This suggests a possibility of an
existence of an analogon of the metrics in the $q$-Minkowski space-time.

Let us define a ``qauntum'' geometry in our
Minkowski space-time by means of a ``quantum metrics''; namely
\begin{equation}
{\rm d}s^2=\pmatrix{{\rm d}x_+,&{\rm d}x_-}\pmatrix{0&1\cr1&0}
\pmatrix{{\rm d}x_+\cr{\rm d}x_-}
\end{equation}
is replaced by
\begin{equation}
{\rm d}s^2_q=\pmatrix{{\rm d}x_+,&{\rm d}x_-}
\pmatrix{0&q^\frac{1}{2}\Gamma\cr q^{-\frac{1}{2}}\Gamma&0}
\pmatrix{{\rm d}x_+\cr{\rm d}x_-},
\end{equation}
where the new generator $\Gamma$ satisfy
\begin{equation}
\Gamma^*=\Gamma,\quad\dot\Gamma=0,\quad\delta(\Gamma)=\sigma^{-2}\otimes\Gamma,
\end{equation}
\begin{eqnarray}
\Gamma m&=&m\Gamma,\\
\Gamma x_\pm&=&q^{\pm1}x_\pm\Gamma.
\end{eqnarray}

The ``line'' element ${\rm d}s^2_q$ can be written in a compact form as
\begin{equation}
{\rm d}s^2_q={\rm d}x^\mu g_{\mu\nu}{\rm d}x^\nu
\end{equation}
with $[g^q_{\mu\nu}]
=\pmatrix{0&q^\frac{1}{2}\Gamma\cr q^{-\frac{1}{2}}\Gamma&0}$, $\mu,\nu=\pm$.
Notice that $[g^{\mu\nu}_q]
=\pmatrix{0&q^\frac{1}{2}\Gamma^{-1}\cr q^{-\frac{1}{2}}\Gamma^{-1}&0}$.

It is easy to see that ${\rm d}s^2_q$ belongs to the center of the space-time
algebra.  Consequently the square of the relativistic momentum reads
\begin{equation}
p^2=p^\mu g_{\mu\nu}p^\nu
=q^\frac{1}{2}p_+\Gamma p_-+q^{-\frac{1}{2}}p_-\Gamma p_+=m^2C^2,
\end{equation}
where $C^2=q^\frac{1}{2}\dot x_+\Gamma\dot x_-
q^{-\frac{1}{2}}+\dot x_-\Gamma\dot x_+$ is the square of the
``light velocity''.  Both $p^2$ and $C^2$ also belong to the center of the
space-time algebra.
Note that the relation of momenta satisfying Eq.~(\ref{qcommut}) and
Eq.~(\ref{momenta}) is of the form $P_\mu=g_{\mu\nu}p^\nu$.

Let us summarize the basic space-time algebra rules
\begin{eqnarray}
x_+x_-&=&qx_-x_+,\\
p_+p_-&=&qp_-p_+,\\
x_\pm p_\pm&=&p_\pm x_\pm,\\
x_\pm p_\mp&=&q^{\pm1}p_\mp x_\pm,\\
\Gamma x_\pm&=&q^{\pm1}x_\pm\Gamma,\\
\Gamma p_\pm&=&q^{\pm1}p_\pm\Gamma.
\end{eqnarray}
It is evident that non-commutativity of coordinates and momenta implies that
there is a number of ``classical'' Heisenbreg-like uncertainity relations.
However, an appriopriate analysis demanded a knowledge of representations of
the above algebra in the (rigged) Hilbert space framework.

\section*{Acknowledgements}
We are grateful to K.~A.~Smoli\'nski for fruitful discussions. One of us (J.R.)
thanks to David Yetter for his kind hospitality.

\end{document}